\documentclass[manuscript]{aastex}
\usepackage{emulateapj5}
\usepackage{graphicx}
\usepackage{times}

\slugcomment{Accepted by ApJL}

\shortauthors{Hopkins, McClure-Griffiths \& Gaensler}
\shorttitle{Evolution of gas and star formation}

\begin{document}

\title{Linked evolution of gas and star formation in galaxies over cosmic history}

\author{A. M. Hopkins\altaffilmark{1},
N. M. McClure-Griffiths\altaffilmark{2},
B. M. Gaensler\altaffilmark{1}
}

\affil{
\begin{enumerate}
\item Institute of Astronomy, School of Physics, The University of Sydney, NSW 2006, Australia; \\
ahopkins@physics.usyd.edu.au
\item Australia Telescope National Facility, CSIRO, P.O. Box 76, Epping NSW
1710, Australia
\end{enumerate}
}

\begin{abstract}
We compare the cosmic evolution of star formation rates in galaxies with
that of their neutral hydrogen densities. We highlight the need for
neutral hydrogen to be continually replenished from a reservoir of
ionized gas to maintain the observed star formation rates in galaxies.
Hydrodynamic simulations indicate that the replenishment may occur naturally
through gas infall, although measured rates of gas infall in nearby galaxies
are insufficient to match consumption. We identify an alternative mechanism for
this replenishment, associated with expanding supershells within galaxies.
Pre-existing ionized gas can cool and recombine efficiently in the walls
of supershells, molecular gas can form {\em in situ\/} in shell walls, and
shells can compress pre-existing molecular clouds to trigger collapse
and star formation. We show that this mechanism provides replenishment rates
sufficient to maintain both the observed HI mass density and the inferred
molecular gas mass density over the redshift range $0\le z\lesssim 5$.
\end{abstract}


\section{Introduction}
\label{int}

Our understanding of the cosmic star formation history (SFH) of galaxies
has progressed significantly over the past decade
\citep[e.g.,][]{Hop:04,HB:06}. In the same time the
space density of neutral hydrogen gas has been measured over the majority
of cosmic history \citep[see Figure~8 of][]{Lah:07}.
The evolution of the atomic hydrogen (HI) in
the universe will be comprehensively determined within the next few years
by extremely sensitive surveys with the next generation of radio telescope
instrumentation \citep[e.g.,][]{vdH:04,Raw:04,Joh:08},
and it is timely to consider mechanisms associated with this evolution.

The space density of HI in galaxies appears to evolve surprisingly little
from $z\approx 5$ to $z\approx 0.2$ \citep{Lah:07}, a span of roughly 10\,Gyr,
the latter half of which sees a decline in the space density of star formation
rate (SFR) in galaxies by almost an order of magnitude \citep[e.g.,][]{HB:06}.
Given the SFR density it is easy to show that the HI plus
molecular gas at high redshift would be exhausted on timescales of a
few Gyr if it were not continually replenished. \citet{Erb:08} presents a
model incorporating gas infall, outflows and consumption by star formation,
to explain both replenishment and the mass-metallicity relation in
high-redshift ($z\approx 2$) galaxies.

Hydrodynamic simulations advocating
hot and cold modes of accretion indicate that the infall rate closely tracks
the star formation rate \citep[e.g.][]{Ker:05,Bir:07}, with star formation
moderated by the rate of infall. The simulations, however, neglect gas
outflows from galaxies, which are a significant component of gas depletion.
The quantitative infall rates predicted are thus insufficient to maintain
a constant HI density in galaxies. Observed rates of gas infall in local
galaxies, also, are only about 10\% of the star formation rate \citep{San:08}.
The difficulties in explaining replenishment through infall leave the
physical mechanism of this replenishment as a critical open question
in galaxy evolution.

In this Letter we suggest a mechanism directly associated with the SFR in
galaxies that can provide the necessary replenishment of neutral gas to
maintain an essentially unevolving, or slowly evolving, HI mass density.
We infer the density of gas required to reproduce the observed SFH in
\S\,\ref{data}. In \S\,\ref{deltarho} we present a number
of models for the replenishment of this gas, and show that a replenishment
proportional to the SFR density can reproduce the necessary gas mass density.
A replenishment mechanism associated with galactic supershells is
detailed in \S\,\ref{disc},
and the results are summarised in \S\,\ref{summ}. Throughout this analysis we
adopt the ``737"\footnote{Thanks to Sandhya Rao \citep{Rao:06} for this
terminology.} cosmology with $H_0=70\,$km\,s$^{-1}$\,Mpc$^{-1}$,
$\Omega_M=0.3$, $\Omega_{\Lambda}=0.7$ \citep[e.g.,][]{Spe:03}.

\section{Estimating the mass density of star forming gas}
\label{data}

While our motivation is to understand the observed lack of significant
evolution in the HI mass density, we approach this by considering the
total mass density of gas available to form stars, which
includes molecular as well as atomic gas. We neglect the intricacies
in the conversion of HI to molecular gas associated with the
star formation process, as this occurs on timescales
very short compared to those involved in this analysis. What is important is
the total reservoir of gas available for star formation, ``star-forming gas,"
$\rho_{\rm SFG}$, at a given redshift, comprised of both atomic and
molecular gas. Star formation is an inefficient process and $\rho_{\rm SFG}$
will likely be somewhat less than the total of the atomic and molecular gas
mass densities.

The cosmic evolution of the HI mass density has been challenging to measure.
There is an implicit assumption that the damped Ly$\alpha$ absorbers used to
measure the HI mass density at $z>0.3$ are representative of the galaxies
used to trace the SFH. The best current measurements suggest that this is
not an unreasonable assumption \citep{ZP:06}, while also highlighting
the difficulty in obtaining observational constraints on $\rho_{\rm SFG}$. In
the absence of direct observational measurements, we estimate $\rho_{\rm SFG}$
indirectly from the observed SFH, using the local relationship between gas
and SFR surface densities from \citet{Ken:98}. We calculate $\rho_{\rm SFG}$
using Equation~(10) of \citet{Hop:05}, and the results are shown in
Figure~\ref{fig:hiandtau}.
The open triangles in Figure~\ref{fig:hiandtau} are derived from the piecewise
linear fit to the SFH of \citet{HB:06}, their Table~2, for their ``SalA"
initial mass function (IMF)\footnote{The SalA IMF is a modified Salpeter IMF
with a turnover below $0.5\,M_{\odot}$ (detailed in \citeauthor{HB:06}
\citeyear{HB:06} and \citeauthor{Bal:03} \citeyear{Bal:03}).}.
The open squares correspond to the SFH of \citet{WTH:08}. This is based on
an evolving IMF, more top-heavy at higher redshift, which reduces the
derived SFR densities and gives lower inferred $\rho_{\rm SFG}$ at high-$z$
than a universal IMF.

These estimates make the implicit assumption that the Kennicutt-Schmidt
law for star formation \citep{Ken:98} is valid at all redshifts, an
assumption questioned by a number of authors \citep{ZP:06,WC:06,WHP:07}.
The relation has, though, been successfully used to explain the
mass-metallicity relationship seen in $z\approx 2$ galaxies by \citet{Erb:08}
and in hydrodynamic simulations by \citet{RK:08} seems to arise as a
natural consequence of gas physics. In the absence of detailed observations
of molecular gas in large numbers of high redshift galaxies,
the assumption that the local relationship holds at higher
redshift is the simplest with which to explore further.

The gas consumption timescale comes from dividing $\rho_{\rm SFG}$ by the
stellar mass density remaining in stars and stellar remnants, inferred from
the SFR density. This gives timescales of $\approx 1-5\,$Gyr with the lower
value at higher redshift where the SFR density is largest. This short
timescale, derived from the global properties of all galaxies, is strikingly
similar to the gas consumption timescales found for individual nearby galaxies
\citep{Won:01,Won:02}.

\section{The evolution of the star forming gas}
\label{deltarho}
To explore the interaction between gas consumption and replenishment we show
the effects of several simple models in Figure~\ref{fig:gasmodels}. All models
begin at a lookback time, $t_{\rm L}$, of $12.55\,$Gyr ($z=6$) assuming
an initial value for $\log(\rho_{\rm SFG}/(M_{\odot}{\rm Mpc}^{-3}))=8.1$,
with $\rho_{\rm SFG}(t_{\rm L})$ calculated simply as the integral over time.
At each time-step gas is consumed by star formation ($-\dot{\rho}_*(t)$) and
a similar amount in gas outflows\footnote{We follow \citet{Erb:08}
in assuming the gas outflow rate from winds exhausts an amount equal
to the SFR, i.e., $-\dot{\rho}_*(t)$. This is a coarse
approximation, but based on galactic outflow rates in the nearby universe
\citep{Vei:05} it may be reasonable when averaged over large populations of
star forming galaxies.}.
Gas is returned to the interstellar medium (ISM) through stellar evolutionary
processes (stellar winds, supernova ejecta) with a recycling fraction
of $R$ \citep[e.g.,][]{Ken:94,Mad:98,Col:01}, adding a factor
$+R\dot{\rho}_*(t)$. For the SalA IMF, $R=0.4$, while other IMFs will have
different recycled fractions \citep[e.g.,][]{HB:06}.
Finally a replenishment factor $K(t)$ is added. This can be expressed as:
\begin{eqnarray}
\label{theeqn}
\rho_{\rm SFG}(t_{\rm L}) & = & \rho_{\rm SFG}(t=12.55) + \nonumber \\
 & &  \int_{t=12.55}^{t=t_{\rm L}} ( - 1.6 \dot{\rho}_*(t) + K(t) )\, dt.
\end{eqnarray}

These simplifications hide a wealth of complex ISM and IGM interactions,
including the fact that material returned to the ISM through stellar
evolution, as well as infalling gas, may contribute to ionized,
or otherwise non-star-forming components as well as directly to
$\rho_{\rm SFG}$. These details are subsumed into the effective
replenishment factor, $K(t)$.

A model with no gas replenishment, $K(t)=0$, is shown in
Figure~\ref{fig:gasmodels} (the dashed line), emphasising the
rapid consumption timescale.
Two models assuming constant rates of replenishment,
$K(t)=0.16\,M_{\odot}\,$Mpc$^{-3}\,$yr$^{-1}$ (dash-dotted line) and
$K(t)=0.21\,M_{\odot}\,$Mpc$^{-3}\,$yr$^{-1}$ (dash-triple-dotted line), are
also shown. The ``bounce" seen in these models arises from an
early excess in consumption followed by progressively decreasing consumption
as $\dot{\rho}_*$ declines for $z\lesssim 1$, predicting excess
$\rho_{\rm SFG}$ at lookback times $t_{\rm L} < 4-6\,$Gyr ($z=0.4-0.7$).

Replenishment factors proportional to the SFR density provide an obvious
way to ensure $\rho_{\rm SFG}$ remains constant with time. A replenishment
factor of $K(t)=1.6\dot{\rho}_*$ in Equation~(\ref{theeqn}) gives
$\Delta \rho_{\rm SFG}=0$ at all redshifts (dotted line in
Figure~\ref{fig:gasmodels}). Different constants of proportionality allow
for slowly varying changes in $\rho_{\rm SFG}$. With $K(t)=1.52\dot{\rho}_*$
\citep[a factor 0.95 of that required for complete replenishment;][]{Erb:08}
the replenishment does not fully balance consumption, and gives a slow
decline in the global gas density (heavy solid line in
Figure~\ref{fig:gasmodels}).

In the following section we consider supershells in the ISM as the driver of a
physical mechanism for replenishment. Supershells are both directly
associated with star formation and are highly efficient at converting
hot phase gas into star-forming gas.

\section{Discussion}
\label{disc}
\subsection{A physical mechanism for replenishment}
\label{physmech}
The ISM in galaxies contains expanding supershells or superbubbles associated
with previous generations of star formation. We propose that the neutral
and molecular gas replenishment in the walls of supershells is sufficient,
and of the appropriate form, to provide a natural mechanism explaining
the relatively flat evolution in the HI mass density.

Supershells are large scale expanding shells of gas driven by supernovae
(SNe) and stellar winds from OB star clusters \citep{Oey:96,OS:98,McCG:01}.
Supershells have long been suggested to have a triggering effect on subsequent
generations of star formation \citep{McC:87,Elm:98,Har:01,Ber:04,Oey:05},
and are effective at replenishing star-forming gas through several
mechanisms. First, supershells are efficient at cooling and recombining
ionized gas through radiative cooling in shell walls \citep{Koo:92}.
This may be critical in converting gas
shock-heated by previous generations of SNe within a galaxy, or new hot-mode
infall gas, to a potentially star-forming state, as the $10^6\,$K gas may
otherwise never cool to support subsequent star formation.
Second, molecular gas can form from neutral gas {\em in situ\/} in shell
walls, where compression and the development of instabilities leads to
sufficiently high neutral gas densities to allow for cooling and
self-shielding on timescales of $10 - 20\,$Myr \citep{Ber:04,Hen:08}.
Finally, they can compress pre-existing molecular material to trigger
molecular cloud collapse and star formation \citep{Elm:98}.
The timescales for these processes are shorter than, or comparable to
the supershell lifetime ($\sim20\,$Myr), which is in turn short
compared to the global gas consumption timescale of several Gyr.

To establish whether the replenishment achievable in supershells is sufficient
to make this mechanism feasible, we first convert the replenishment rates
given in \S\,\ref{deltarho} into a replenished mass per SN event. A
replenishment rate proportional to the SFR density is also proportional to
the rate of supernova type II (SNII\footnote{Here and throughout we assume
the inclusion of all core-collapse supernovae: types II, Ib, and Ic}).
Converting a proportionality to SFR density into one depending on
the SNII rate, $\dot{\rho}_{\rm SNII}$, depends on the assumed IMF. From
\citet{HB:06} $\dot{\rho}_{\rm SNII}=(0.00915/M_{\odot})\,\dot{\rho}_*$ for
the SalA IMF. The replenishment rate $K(t)=1.6\dot{\rho}_*$ becomes
$K(t)=174.9\dot{\rho}_{\rm SNII}\,M_{\odot}$.

The other extreme choice of IMF consistent with the normalization
of the SFH \citep{HB:06} is that
of \citet[hereafter the BG IMF]{Bal:03}. For the BG IMF
$\dot{\rho}_{\rm SNII}=(0.0132/M_{\odot})\,\dot{\rho}_*$. The recycled
fraction is $R=0.56$ \citep{HB:06}, changing the consumption term
in Equation~(\ref{theeqn}) to $-1.44\dot{\rho}_*$. The corresponding
replenishment rate is $K(t)=109.1\dot{\rho}_{\rm SNII}\,M_{\odot}$.
These extremes imply that, depending on the IMF, sufficient gas
replenishment to maintain a constant HI mass density with redshift
would be achieved if each SN event caused the recombination and
cooling of $\approx 110-180\,M_{\odot}$ of gas. These IMFs are the
extrema given the SFH normalization limits, and most reasonable IMFs
should result in masses within this range.

Detailed measurements to confirm molecular gas formation within supershells
are observationally challenging. We use the limited data currently
available to assess the replenishment rates associated with supershells,
and to establish whether at least one well-studied supershell achieves the
required rate.

\citet{McCG:05} and \citet{Daw:08} have shown explicit cases of molecular
clumps along the edges of supershells, suggestive of some degree of
{\em in situ\/} formation, with a significant amount of molecular material
associated with the supershell walls. The supershell investigated by
\citet{Daw:08} is associated with about $2\times10^5\,M_{\odot}$ of molecular
gas, of which those authors estimate that $80\%$ likely comes from a
pre-existing giant molecular cloud. Of the remaining
$\approx 4\times10^4\,M_{\odot}$ of molecular gas it is difficult to determine
how much is pre-existing and how much has been cooled and recombined by
the expansion of the shell. We can use $\approx 4\times10^4\,M_{\odot}$ as an
upper limit to the replenishment rate. About 30 stars with stellar
mass $M_*> 7\,M_{\odot}$ are required to form this supershell, including stars
that may not yet have gone supernova. This gives
$\lesssim 1300 - 2000\,M_{\odot}$ of molecular mass replenished per SN event,
a limit comfortably encompassing the required rate. This upper limit 
could change significantly depending on the fraction of pre-existing molecular
material and also on the fraction of stars that have not yet gone supernova.

Not all SNe lie within supershells, although \citet{HL:05} estimate that
a minimum of $65\%$ of SNII should occur in superbubbles, increasing to
$\approx 80-90\%$ when the spatial and temporal correlations of stellar
clusters are considered. If $80\%$ of SNII are associated with supershells,
for example, this would increase the required replenishment rate per SN
to $\approx 140-230\,M_{\odot}$. But even if as few as 10\% of all SNII
contribute in this way to the replenishment,
the rate implied by the results of \citet{Daw:08} would still be sufficient.
This confirms that the necessary replenishment rates are likely to be
achievable within supershells.

The observed decline by a factor of two in the HI mass density may
be a natural consequence of a replenishment rate about 95\% of that
required to match consumption, as shown by the heavy solid line in
Figure~\ref{fig:gasmodels}. If the actual replenishment rate from
supershells lies somewhere between the required rate and our derived
upper limit, though, there may in fact be too much newly replenished gas
to allow any decline in the neutral gas mass density. A possible resolution
in this scenario would be increasing the proportionality between
the gas outflow rates and the SFR as redshift decreases. This is not
unreasonable, as the SFH is becoming progressively more dominated by
lower-mass galaxies with decreasing redshift \citep{Jun:05,Pan:07,Mob:08}.
Galaxies with stellar masses $M_* \lesssim 10^{10}\,M_{\odot}$ dominate the
SFH at $z\lesssim 1$ \citep{Mob:08}. Such low-mass galaxies lose more mass
in gas outflows in proportion to their SFR than high-mass galaxies, simply
due to the former's shallower potential wells
\citep[e.g.,][]{DS:86,MF:99,FT:00}. This effect may contribute to
the slow decline in the HI mass density.

\subsection{Limitations of the proposed mechanism}
\label{limitations}
We have treated a number of complex physical processes in very general terms.
While being cautious of oversimplification, we have attempted to capture the
essential interactions between star formation, recycling from stellar
evolutionary processes, ISM processes of heating and ionization,
recombination, cooling and molecule formation, together with infall from
the IGM, and outflow of ISM material. Most of this complexity is
concealed within the replenishment factor $K(t)$.

One issue is that stellar winds and SNe contribute to all
components of the ISM rather than solely to $\rho_{\rm SFG}$.
In a ``galactic fountain" \citep{SF:76,HB:90},
infalling gas will contribute to, and outflowing gas will strip from,
all components. If recycled gas includes a component that never
subsequently forms stars (such as some recycled
gas in the ionized phase being ejected from the galaxy before contributing
to star formation), the factor $+R\dot{\rho}_*(t)$ in Equation~(\ref{theeqn})
will be reduced, and $K(t)$ will need to be increased to compensate.

Our quantitative results strongly depend on the assumed gas outflow rate.
Variations by a factor of two or so in either direction will still result
in a constant or slowly varying HI mass density, as long as a proportionality
with the SFR of the host galaxies remains \citep[as suggested by][]{Vei:05}.
The chosen outflow rate is an effective average over all star forming galaxies,
and is consistent with observed trends \citep[e.g.,][]{Mar:99,Pet:00,Vei:05}.
While individual galaxies show a large observed scatter between outflow
rates and SFRs, for the ensemble properties of the total population
this assumption should be robust.

The proposed replenishment through the supershell mechanism is not
inconsistent with some simultaneous replenishment through infall.
Metallicity considerations, which we do not address here, do require
infall of some low metallicity gas \citep{Erb:08}, and gas infall in local
galaxies is well established \citep[e.g.,][]{Bla:07,San:08}, although the
observed infall rate is insufficient to match consumption.

\section{Summary}
\label{summ}
We have identified a possible resolution to the puzzle of why the HI mass
density of the universe evolves so little for so much of cosmic history.
We propose that replenishment is driven by supershells associated with
star forming complexes in galaxies. Pre-existing ionized
gas efficiently cools and recombines in supershell walls.
Molecular gas forms {\em in situ\/} in shell walls, and molecular
material is compressed to trigger cloud collapse and star formation.
This mechanism provides a natural explanation
for replenishment that has the desired proportionality to the SN rate.
The level of replenishment observed in a Galactic supershell \citep{Daw:08}
appears more than sufficient to provide the required replenishment
rate of $\approx 110-180\,M_{\odot}$ per SN event.

The factor of two decline in the HI density between $z\approx 0.2$
and $z=0$ could be explained through either (1) a replenishment rate that is
marginally lower than that required to exactly balance gas consumption;
or (2) the inability of low-mass galaxies, which dominate the star formation
history in this epoch, to retain their newly formed HI; or, perhaps more
likely, a combination of both.

\acknowledgements
We thank the referee for their comments and helpful input on hydrodynamic
simulations. We thank Joss Bland-Hawthorn and Leo Blitz for
valuable discussions that helped to improve this analysis.
AMH acknowledges support provided by the Australian Research Council (ARC)
in the form of a QEII Fellowship (DP0557850). BMG acknowledges support
from the ARC through a Federation Fellowship (FF0561298).

\begin{figure*}
\centerline{\rotatebox{-90}{\includegraphics[width=10.0cm]{rhogastime_new3.ps}}}
\caption{The cosmic history of neutral and ``star-forming" gas mass density.
Solid circles: The HI density as shown in Figure~8 of \citet{Lah:07}.
From high to low redshift the
data come from \citet{Pro:05} at $t_{\rm L} \gtrsim 10\,$Gyr ($z> 1.9$);
from \citet{Rao:06} between $5\lesssim t_{\rm L} \lesssim 9\,$Gyr,
($0.61 \le z \le 1.22$); the stacking measurement from \citet{Lah:07}
at $t_{\rm L}=2.8\,$Gyr, ($z=0.24$); and the HIPASS measurement of
the local HI density from \citet{Zwa:05}.
Open red triangles: $\rho_{\rm SFG}$, the neutral plus molecular gas density
inferred from the SFH from \citet{HB:06}, assuming the
Kennicutt-Schmidt relation for star formation;
Open blue squares: $\rho_{\rm SFG}$ as
inferred from the SFH of \citet{WTH:08}, which assumes an evolving initial
mass function. The solid magenta and cyan lines are nominal parameterisations
of the possible evolution of the total gas reservoir for each of the SFH cases.
 \label{fig:hiandtau}}
\end{figure*}

\begin{figure*}
\centerline{\rotatebox{-90}{\includegraphics[width=10.0cm]{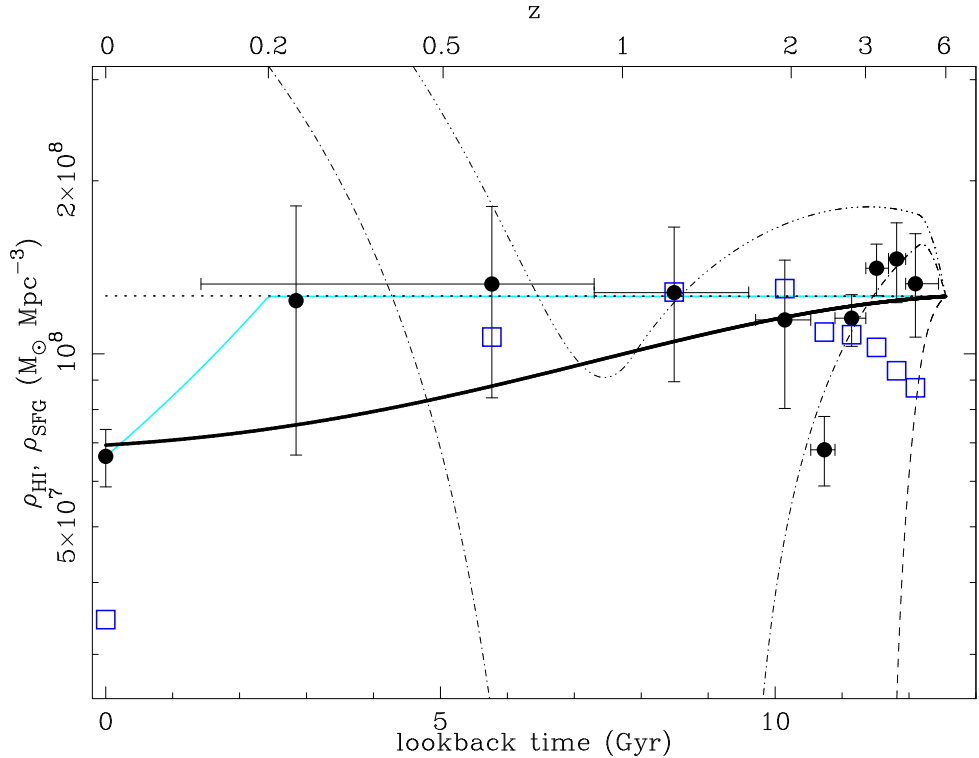}}}
\caption{Models for the evolution in the gas reservoir. The HI mass density
(solid circles), $\rho_{\rm SFG}$ (open blue squares), and solid cyan line
are as in Figure~\ref{fig:hiandtau}. Dashed line: Predicted gas
reservoir evolution assuming no replenishment; Dash-dot and dash-triple-dot
lines: Two different rates of constant replenishment;
Dotted line: Replenishment rate proportional to the SN rate;
Heavy solid line: Replenishment rate proportional to
the SN rate, but at a factor of 0.95 of that required to balance
the consumption. See text for further details.
 \label{fig:gasmodels}}
\end{figure*}

\end{document}